\documentclass[aps,prb,reprint,superscriptaddress]{revtex4-1}

\pdfoutput=1

\usepackage{graphicx}
\graphicspath{{}}

\newcommand{\e}{\mathrm{e}}

\begin{document}


\title{Noncentrosymmetric plasmon modes and giant terahertz photocurrent in a two-dimensional plasmonic crystal}


\author{V.\,V.\, Popov}
\affiliation{Kotelnikov Institute of Radio Engineering and Electronics, Russian Academy of Sciences,
Saratov 410019, Russia}
\affiliation{Saratov State University, Saratov 410012, Russia }
\affiliation{Saratov Scientific Center of the Russian Academy of Sciences, Saratov 410028, Russia}
\author{D.\,V.\, Fateev}\email{FateevDV@yandex.ru}
\affiliation{Kotelnikov Institute of Radio Engineering and Electronics, Russian Academy of Sciences,
Saratov 410019, Russia}
\author{E.\,L.\,Ivchenko}
\affiliation{Ioffe Physical-Technical Institute of the Russian Academy of Sciences, 
St.~Petersburg 194021, Russia}
\author{S.\,D.\,Ganichev}
\affiliation{Terahertz Center, University of Regensburg, 93040, Regensburg, Germany}


\date{\today}

\begin{abstract}
We introduce and theoretically study the plasmon-photogalvanic effect in the planar noncentrosymmetric plasmonic crystal containing a homogeneous two-dimensional electron system gated by a periodic metal grating with an asymmetric unit cell. The plasmon-photogalvanic DC current arises due to the two-dimensional electron drag by the noncentrosymmetric plasmon modes excited under normal incidence of terahertz radiation. We show that the collective plasmon modes of the planar plasmonic crystal become strongly noncentrosymmetric in the weak coupling regime of their anticrossing. Large plasmon wavevector (which is typically by two-three orders of magnitude greater than the terahertz photon wavevector) along with strong near-field enhancement at the plasmon resonance make the plasmonic drag a much stronger effect compared to the photon drag observed in conventional two-dimensional electron systems.
\end{abstract}

\pacs{73.20.Mf, 78.67.Pt, 07.57.-c}

\maketitle

\section{Introduction}
In natural crystals lacking the center of symmetry, a DC electric current can be generated under the action of optical illumination. This phenomenon is commonly referred to as the photogalvanic effect. \cite{1IvchSL} Physically, the DC photogalvanic signal appears as a result of nonlinear mixing of the optical fields with frequencies $\omega$ and $-\omega$ in a noncentrosymmetric nonlinear crystal. Generally, the absence of inversion symmetry in a natural material yields the asymmetry in the electronic response while the optical field may be symmetric or even homogeneous. \cite{2GanTE,3IvchOS} Artificial noncentrosymmetric microperiodic structures with a two-dimensional (2D) electron system can also exhibit photogalvanic DC responses due to the asymmetry of the electronic response. \cite{4Chep,5Weber,Olbrich2009,Ivchenko2011,26Olb,27Nali} Under oblique incidence of the electromagnetic wave, the photocurrent can be excited even in a centrosymmetric homogeneous 2D electron system, including graphene, due to the photon drag effect. \cite{5Weber,6Glaz} In this case, the asymmetry of the incident electric field in the plane of the 2D electron system lifts the symmetry constrains for exciting a non-zero photocurrent.

It is also known that a photocurrent can be induced on the surface of a centrosymmetric metal material if the surface of metal is patterned in the form of diffraction grating with an asymmetric unit cell.  \cite{7Hatan} The metal grating couples light to the surface plasmon-polaritons on the metal surface since the reciprocal wavevectors of the periodic grating ensure the momentum conservation between the incident light and the surface plasmon-polaritons.  In case of the diffraction grating with a broken symmetry, \cite{8Bonod,9Christ,10Bai,11Rosz,12Wood} the excited plasmon-polaritons are predominantly unidirectional and can effectively drag the free carriers which results in a net photocurrent. \cite{7Hatan} 

\begin{figure}[ht]
\includegraphics{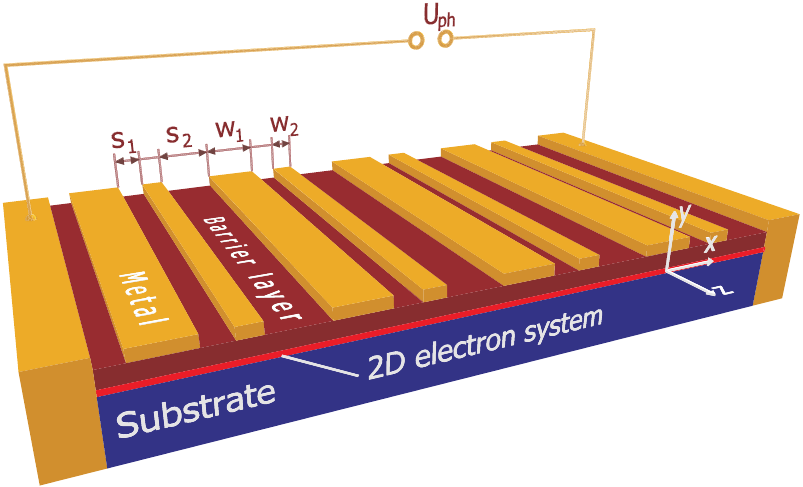}
\caption{Schematic view of the 2D-PlCr with an asymmetric unit cell. External electromagnetic wave is incident  from the top upon the 2D-PlCr at normal direction to its plane  with the polarization of the electric field across the metal grating strips. Terahertz photocurrent generates the photovoltage $U_{\rm{ph}}$ between the open side contacts.}
\label{fateev1}
\end{figure}

While the surface plasmon-polaritons in metals are excited in visible and near-infrared frequency ranges, the frequencies of the plasmons in 2D electron systems, \cite{13PopJInfr} including graphene, \cite{14Kopp} fall into the terahertz (THz) range. In contrast to the surface plasmon-polaritons in metals, the wavelength of the plasmons in the 2D electron systems is much shorter (by two and even three orders of magnitude) than the wavelength of electromagnetic radiation of the same frequency. Therefore, the THz radiation can be coupled to the 2D plasmons by using the grating coupler with the (sub)micron period. \cite{13PopJInfr,15Geng,16Gao} The two-dimensional electron system with a periodic grating can be viewed as a 2D plasmonic crystal (2D-PlCr) since, in such a system, the electromagnetic near-field has a plasmonic nature and its wavelength is determined by the periodicity of the grating. The photocurrent induced by THz radiation in 2D-PlCr originates from the ponderomotive (plasmonic) nonlinearities of the free-electron motion in a 2D electron system \cite{13PopJInfr,17DyaSh,18Ivch14} where the centrosymmetry is broken by a DC current passing through 2D electron system \cite{19Aiz06,20Aiz07} or because of geometrical asymmetry of the grating coupler unit cell \cite{21PopAPL11} (see Fig. \ref{fateev1}).
 In the 2D-PlCr, the asymmetry of the plasmonic near-field in the structure unit cell is essential for an appearance of the plasmonic THz photocurrent while the electronic response in the 2D electron system can be symmetric or even homogeneous. The strength of the plasmonic near-field can be considerably enhanced by exciting the plasmon resonances in 2D-PlCr. As demonstrated both theoretically \cite{21PopAPL11} and experimentally, \cite{22Wata,23Kurita,24Step} 2D-PlCr with a geometrically asymmetric unit cell can exhibit a giant photogalvanic response at the THz frequencies.

As shown in Ref.~\onlinecite{25PopAPL13}, the plasmonic photoresponse of the 2D-PlCr with an asymmetric unit cell is caused by two different physical mechanisms (which might be comparable in strength), namely, the plasmonic drag and plasmonic ratchet effects. Both mechanisms rely on an asymmetric profile of the plasmonic field in the 2D-PlCr unit cell. Plasmonic ratchet effect is possible only in 2D-PlCr with a spatially periodic 2D electron system and can exist also for a centrosymmetric profile of the plasmonic field (which should however be shifted relative to the profile of the spatial modulation of 2D electron system though). Although the plasmonic drag effect can exist even in a homogeneous 2D electron system, the noncentrosymmetry of the plasmonic field in 2D-PlCr is required in this case. In general, the asymmetric profile of the plasmonic field is a direct consequence of the geometrically asymmetric unit cell. However, the relation between the asymmetry of the 2D-PlCr unit cell and that of the plasmonic-field profile is intricate, nontrivial and dependent on a particular kind of 2D-PlCr. Although the plasmonic field is expected to increase at the plasmon resonance frequency, the excitation of the plasmon resonance does not in general guarantee a resonant increase in the asymmetry of the plasmonic field. Therefore, it has not been fully understood up to now why the asymmetry of the plasmonic field in 2D-PlCr is so strong to cause the ultra-high THz plasmonic photoresponse predicted in Ref. \onlinecite{21PopAPL11} and experimentally observed in Refs. \onlinecite{22Wata,23Kurita,24Step}.

In this paper, we consider the 2D-PlCr formed by a homogeneous 2D electron system screened by a periodic metal grating with an asymmetric unit cell and focus on the noncentrosymmetric differential plasmonic drag effect induced in such a system by the THz wave incident at normal direction upon the 2D-PlCr plane. The metal grating is formed by two coplanar interdigitated sub-gratings having the same period but different widths of the metal strips $w_1$ and $w_2$. Geometrical asymmetry of the unit cell can be varied by shifting one sub-grating relative to the other in the grating plane, Fig. \ref{fateev1}. We show that the plasmon resonances with strong asymmetry of the near electric field can be excited in the anticrossing regime between different plasmon modes of the 2D-PlCr which leads to plasmon-photogalvanic current caused by the  differential plasmon drag. In Sec. II, we obtain the formula for calculating the differential plasmon-drag current in a homogeneous 2D electron system. Plasmon mode excitation in the 2D-PlCr with a symmetric unit cell is considered in Sec. III. Noncentrosymmetric plasmon modes and induced plasmon-photogalvanic current in 2D-PlCr with an asymmetric unit cell are studied in Sec. IV. We summarize the main results in Sec. V.

\section{NONCENTROSYMMETRIC PLASMON DRAG. Numerical details}
We assume that the external electromagnetic THz wave with the electric field
$$E_x^{({\rm ext})}(z,t) = E_0 \exp [-{\rm i}\omega (t + z/c)]$$
polarized across the metal grating strips (the $x$-direction), is incident normally from vacuum upon the grating plane.
Here $\omega$ is the wave frequency, and $c$ is the speed of light. The DC photocurrent density generated due to the ponderomotive
nonlinearities in a homogeneous 2D electron system by normally incident THz wave can be described by the equation \cite{18Ivch14}
\begin{equation} \label{jx}
j_x = - \frac{1}{L}\frac{e \tau}{2m\omega}\int\limits_{- L/2}^{L/2} {\rm Im} \left\{ \sigma(\omega) \tilde{E}^*_x \frac{\partial}{\partial x} \tilde{E}_x \right\}dx,
\end{equation}
where $\sigma(\omega) = e^2 N_{2D} \tau/[m(1 - {\rm i} \omega \tau)]$  is the dynamic Drude conductivity of 2D electron system with $-e$
and $m$ being the electron charge ($e > 0$) and effective mass, $N_{2D}$ is the sheet electron density, $\tau$ is the
characteristic electron scattering time, and $ L= w_1+s_1+w_2+s_2$ is the grating-gate period. The real-valued electric field in the plane of 2D electron system is defined as
$E_x(x,t) = [ \tilde{E}_x(x) \exp(-{\rm i}\omega t)+\rm{c.c.}]/2$. For open side contacts (see Fig. \ref{fateev1}), the photocurrent $j_x$ generates the photovoltage $U_{\rm{ph}}=W R j_x$ between the side contacts, where $R$ and $W$ are the DC resistance and width (across the photocurrent) of the 2D electron system, respectively. One can rewrite Eq. (1) in the Fourier representation as
\begin{equation} \label{jxsum}
j_x = - \frac{e \tau}{2m\omega} {\rm Re} [ \sigma(\omega) ] \sum\limits_{p > 0} q_p \left( | \tilde{E}_{x,p} |^2 - | \tilde{E}_{x,-p} |^2 \right),
\end{equation}
where $q_p = 2p\pi/L$ and
\begin{equation} \label{fourier}
\tilde{E}_{x,p} =  \frac{1}{L}\int\limits_{- L/2}^{L/2} \tilde{E}_x(x) \e^{- {\rm i} q_p x} dx \hspace{3 mm} (p=0, \pm 1, \pm 2 \dots)
\end{equation}
are the amplitudes of the Fourier harmonics of the complex-valued electric field $\tilde{E}_x(x)$  in the 2D electron system. It is seen
from Eq. (2) that the net DC photocurrent results from the plasmonic drag \cite{18Ivch14,25PopAPL13} by the oppositely travelling Fourier harmonics of
the near electric field in 2D electron system. The non-zero differential plasmon drag current appears only for a noncentrosymmetric near field in the 2D
electron system where $|\tilde{E}_{x,p} |^2 \neq | \tilde{E}_{x,-p} |^2$. The sign of the photocurrent is determined solely by the
asymmetry of the near electric field because all other quantities entering Eq. (2) are positive and it reverses under the mirror reflection of the
near field $|\tilde{E}_{x,p} |^2 \to | \tilde{E}_{x,-p} |^2$. Therefore, this effect can be termed the plasmon-photogalvanic effect because it is totally determined by the noncentrosymmetry of the near electric field in the 2D electron system. It is worth noting that the noncentrosymmetric plasmonic drag is allowed even for the normal incidence of THz wave upon the 2D-PlCr plane despite the lack of the in-plane component of the photon momentum in the incoming THz wave. This becomes possible due to an asymmetric profile of the plasmonic field with non-zero in-plane component of the plasmon momentum.

We simulate the interaction of the external electromagnetic wave with the 2D-PlCr shown in Fig. 1 within the self-consistent electromagnetic approach based on the integral equation method described in Ref. \onlinecite{28fateev}. We use the material parameters $N_{2D}=7.5\times10^{12}$~cm$^{-2}$  and $m = 0.2m_0$, where $m_0$ is the free electron mass, which are characteristic for the 2D electron system in AlGaN/GaN heterostructures. \cite{29Mura} The dielectric constant of the structure constituent materials is assumed to be $\varepsilon$ = 9.5 for each layer of the heterostructure. The metal grating is separated from the 2D electron system by a barrier layer of thickness $d = 30$~nm. In our simulations, we assume the metal grating strips being perfectly conductive and infinitely thin. This is a quite justified and commonly used assumption at THz (and lower) frequencies where the metals are characterized by a high real conductivity and hence the electromagnetic field penetrates into the metal only by a short skin depth (which is much shorter than the electromagnetic wavelength).

In 2D-PlCr under consideration, there are both screened and unscreened parts of 2D electron system. Accordingly, the screened and unscreened plasma oscillations can be, in principle, excited in different parts of the 2D-PlCr. However, the screened and unscreened plasmons have different dispersion characteristics (typically, the frequencies of the unscreened plasmons are much higher as compared to the screened plasmon frequencies for the same plasmon wavevector) so that the screened and unscreened plasmon resonances can be excited independently in 2D-PlCr. For definiteness, we limit ourselves to studying the screened plasmon resonances excited at low THz frequencies. The dispersion relation for plasmons in 2D electron system screened by a perfectly conductive plane is \cite{13PopJInfr}
\begin{equation} \label{frequency}
\omega= \sqrt{ \frac{2 \pi e^2 N_{2D} k}{ m \bar{ \varepsilon } } }\:,
\end{equation}
where $k$ is the plasmon wavevector and  $\bar{\varepsilon} = \varepsilon [1 + {\rm coth} (kd)]$. For $kd \ll 1$, one has ${\rm coth} (kd) \approx (kd)^{-1}$  and the dispersion relation Eq. (3) is reduced to
\begin{equation} \label{frequency2}
\omega= k \sqrt{ \frac{2 \pi e^2 N_{2D} d}{ m \varepsilon} }\:.
\end{equation}

\section{Plasmon modes in the structure with a symmetric unit cell }
\begin{figure}[ht]
\includegraphics{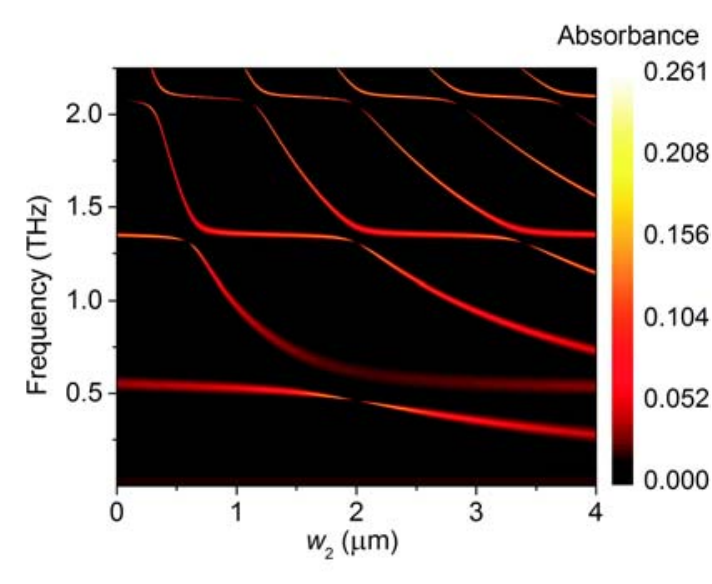}
\caption{(Color online) Plasmon absorption spectrum of the 2D-PlCr with a symmetric unit cell ($s_1 = s_2 = 1.5~\mu$m) depending on the variation of the strip width $w_2$ for a fixed strip width $w_1 = 2~\mu$m and the electron scattering time $\tau=7\times10^{-11}$~s.}
\label{fateev2}
\end{figure}

In order to reveal the characteristic properties of the plasmonic spectrum in a 2D-PlCr, we assume quite a long 2D electron scattering time $\tau=7 \times 10^{-11}$~s and start with considering the 2D-PlCr with a symmetric unit cell. In this case the strips of one sub-grating are exactly centered in the middle of intervals between the adjacent strips of the other sub-grating, $s_1 = s_2 \equiv s$. Figure \ref{fateev2} shows the plasmon absorption spectrum for the symmetric structure as a function of the strip width $w_2$ for fixed values of $w_1 = 2~\mu$m and $s = 1.5~\mu$m.

In the simplest single-strip approximation we can consider the plasmon modes as being excited only in the area under an individual metal strip of the width $w_l$ $(l = 1,2)$. Taking the open circuit boundary conditions at the strip sides we obtain the plasma resonant frequencies $\omega(l,n_l)$ as given by Eqs. (4) and (5) for the size-quantized plasmon wavevector
$$
k_l(n_l) = \pi n_l/w_l\hspace{3 mm}  (n_l = 1, 2, 3\dots)\:.
$$
Therefore, in this lowest-order approximation two types of screened plasmon modes with $l=1$ and $l=2$ are excited in 2D-PlCr and the Bloch states are constructed by the linear combinations of the single-strip excitations with the same $l$ and $n_l$. Under the normal incidence of THz radiation only the even modes (with odd indexes $n_l$) are excited. (Note that the profile of the in-plane electric field of the even plasmon mode in 2D electron system across a grating-gate strip is even while that of the  electron density distribution is odd relative to the strip center.)  In the next order approximation, a coupling between nearest neighbors is switched on. The resonant coupling between the two different plasmon modes becomes important if frequencies, $\omega_1(n_1)$ and $\omega_2(n_2)$ with $n_1$ and $n_2$ of the same parity approach each other.

\begin{figure}[ht]
\includegraphics{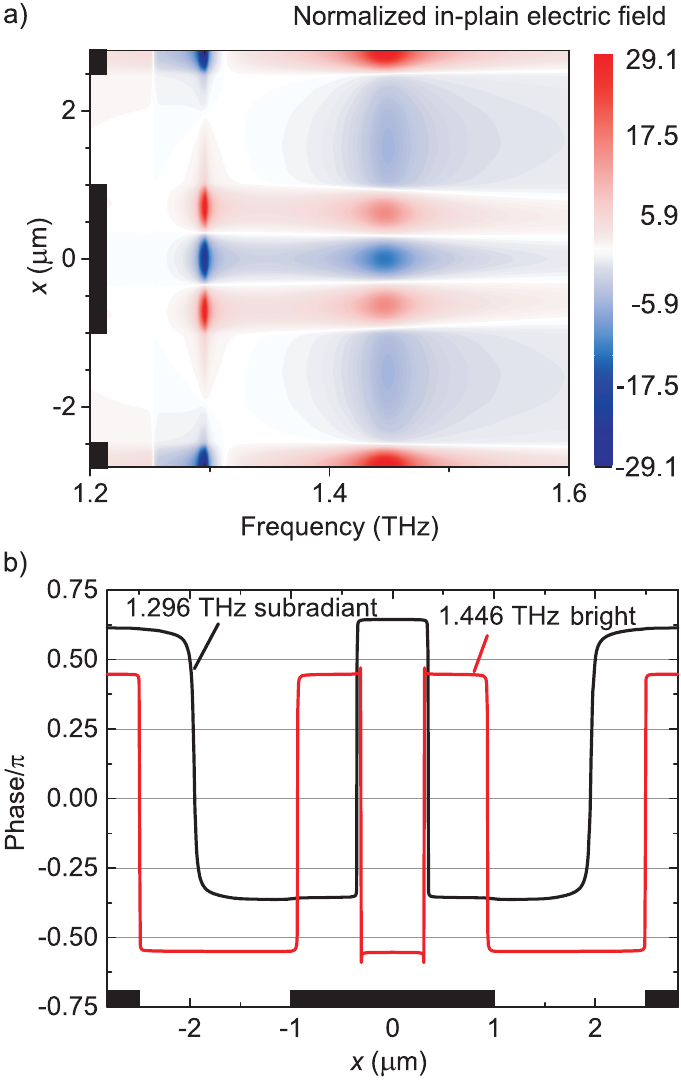}
\caption{(Color online) (a) Normalized in-plane electric field in the 2D electron system of the 2D-PlCr with a symmetric unit cell at the anticrossing regime between the $n_1 = 3$ and $n_2 = 1$ plasmon modes at frequency around 1.35 THz for $w_2 = 628$~nm and $\tau=7\times10^{-11}$~s. The coupled plasmon mode resonating at frequency 1.446 THz is bright while the coupled plasmon mode excited at frequency 1.296 THz is subradiant. The in-plane electric field in 2D electron system is normalized to the electric-field amplitude of the incident THz wave.  (b) The phases of the electric field of the two different coupled plasmon modes as functions of the $x$-coordinate within the 2D-PlCr unit cell. The difference between the electric fields oscillating at different antinodes of the plasmon modes equals $\pi$. The positions of the metal-grating strips are shown by thick black bars along the $x$-coordinate.}
\label{fateev3}
\end{figure}

The horizontal lines outside the narrow anticrossing regions in Fig. \ref{fateev2} represent the frequencies of the plasmon modes excited under conductive strips of the width $w_1$ while the frequencies of the plasmon modes excited under conductive strips of the width $w_2$ are inversely proportional to $w_2$ according to the dispersion law (4). The anticrossing behavior between the plasmon modes excited under the conductive strips of the two different sub-gratings is clearly seen in Fig. \ref{fateev2}. As followed from the dispersion relation (4), the frequencies of two plasmon modes coincide if the widths $w_1$ and $w_2$ are commensurable, i.e.,
if $w_1/w_2 = n_1/n_2$. At the crossing points, the degeneracy of the modes is lifted due to their interaction giving rise to the plasmon mode splitting and the formation of two coupled modes with the frequencies
\[
\omega_{\pm} = \frac{\omega_1(n_1) + \omega_2(n_2)}{2} \pm \sqrt{ \left[ \frac{\omega_1(n_1) - \omega_2(n_2)}{2} \right]^2 + t_{12}^2}\:,
\]
where $t_{12}$ is the effective coupling constant equal to the sum of the coupling constants of the plasmon mode in a particular strip with the plasmon modes in its left- and right-hand side neighbor  strips. For one of the coupled modes the net dipole moments in the areas under all the metal strips have the same polarity whereas those for the other coupled mode have opposite polarities under the adjacent conductive strips, Fig. \ref{fateev3}(a).
Correspondingly, the former mode is bright and the latter is subradiant and can become optically inactive (dark) at some frequency because of the destructive interference of the dipole moments oscillating under the strips of the different sub-gratings. In Fig. \ref{fateev2}, the dark coupled mode is seen as a blind spot in the low-frequency branch of the split modes.

The plasmon mode profiles of the resonances shown in Fig. \ref{fateev2} are symmetric with respect to the vertical plane of mirror symmetry of the 2D-PlCr, see Fig. \ref{fateev3}(a). Because the condition $\tilde{E}_{x,p} = \tilde{E}_{x,-p}$ is satisfied in this case for both bright and subradiant plasmon modes, no plasmon-photogalvanic current flows in 2D-PLCr according to Eq. (2).

\begin{figure}[ht]
\includegraphics{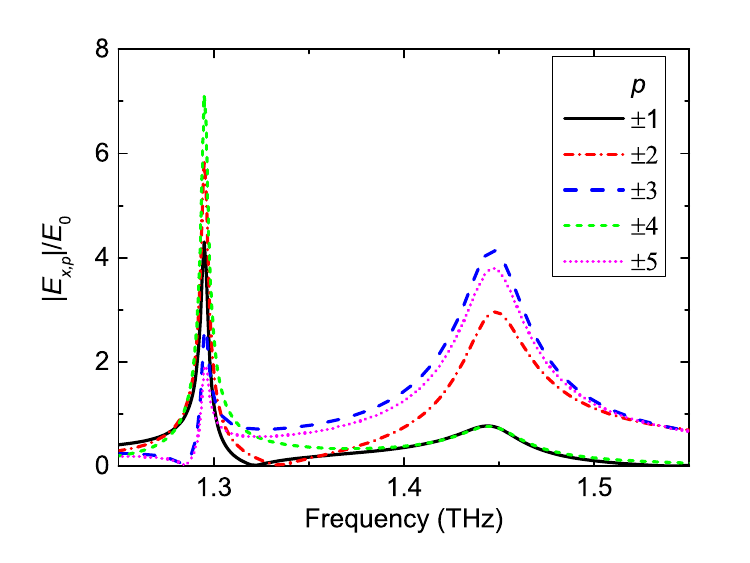}
\caption{(Color online) The spectra of different spatial Fourier harmonics of the in-plane component of the plasmonic electric field in the 2D electron system  near the anticrossing regime at frequencies around 1.35 THz for $w_2 = 628$~nm and $\tau=7\times10^{-11}$~s in the 2D-PlCr with a symmetric unit cell. The amplitudes of the Fourier harmonics are normalized to the electric-field amplitude in the incident THz wave.}
\label{fateev4}
\end{figure}

Figure \ref{fateev4} shows the spectra of the spatial Fourier harmonics of the plasmonic electric field $\tilde{E}_x(x)$ near the anticrossing point at frequencies around 1.35~THz for the 2D-PlCr with a symmetric unit cell. It is seen that different Fourier harmonics resonate stronger in different coupled resonances due to different plasmon mode profiles in those resonances. The resonance linewidth is defined by the electron scattering in the 2D electron system (with the contribution $1/\tau$ into the linewidth of any plasmon resonance) and radiative damping of the plasmon mode. The radiative damping is considerably smaller for the subradiant plasmon modes and, accordingly, the linewidth of the low-frequency resonance is several times narrower than that of the high-frequency (bright) one. Intriguing feature, however, is that the intensity of the absorption resonance of the subradiant plasmon mode can be even higher than that of the bright plasmon mode at some frequency close to the blind spot in the subradiant plasmon-mode branch. This is a consequence of the impedance matching requirements called the equipartition condition: the strongest absorption occurs when the  dissipative and radiative contributions to the resonance linewidth are equal. \cite{PopovPeralJAP} The radiative damping of the subradiant plasmon mode gradually decreases when approaching the blind spot. Hence, at some point of the dispersion curve close to the blind spot, the radiative damping of the subradiant mode becomes equal  to a relatively small dissipative damping, $1/ \tau$, which causes the strongest absorption at the subradiant plasmon-mode resonance at this point.    

\section{Plasmon modes in the structure with an asymmetric grating. PLASMON-PHOTOGALVANIC CURRENT}

Now we turn to the 2D-PlCr with asymmetric unit cells where the conductive strips of width $w_2$ are shifted from the center of the gap between the conductive strips of width $w_1$ making the inter-strip spacings $s_1$ and $s_2$ different. For certainty, we take $s_1$ = 1~$\mu$m and $s_2$ = 2~$\mu$m. As mentioned earlier, a geometrical asymmetry itself does not guarantee strong asymmetry of the near field in the structure. For example, the Fourier harmonics of the electric near field of a bare metal grating (without 2D electron system at all) with an asymmetric unit cell are highly symmetric. \cite{30IvchPet14} Consequently, the photoexcitation of a plasmon resonance in the 2D electron system with an asymmetric unit cell, in general, may not lead to  an enhancement of the plasmonic field asymmetry if, e.g., the plasmon mode is excited only under conductive strips of one sub-grating. It follows then that only the excitation of coupled plasmon modes resonating under the conductive strips of both sub-gratings, together with a geometrical asymmetry of the 2D-PlCr unit cell, could resonantly enhance the asymmetry of the plasmonic field in the 2D-PlCr.

\begin{figure}[ht]
\includegraphics{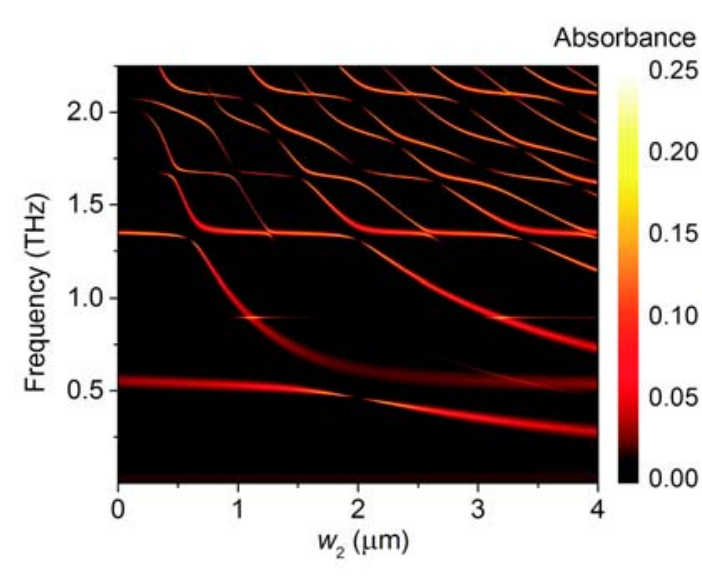}
\caption{(Color online) Plasmon absorption spectrum of the 2D-PlCr with an asymmetric unit cell ($s_1 = 1~\mu$m, $s_2 = 2~\mu$m) depending on the strip width $w_2$ for a fixed strip width $w_1 = 2~\mu$m and the electron scattering time $\tau=7\times10^{-11}$~s.}
\label{fateev5}
\end{figure}

Figure \ref{fateev5} shows the THz absorption spectrum of the 2D-PlCr with an asymmetric unit cell depending on variation of the strip width $w_2$ for a fixed width $w_1$. Although this spectrum resembles that in the 2D-PlCr with a symmetric unit cell shown in Fig. \ref{fateev2}, it is much richer.
Apart from the even plasmon modes (with odd $n_l$), the odd plasmon modes can also be excited under conductive strips of either sub-grating due to the asymmetry of the 2D-PlCr unit cell. The odd modes $\omega_1(n_2)$ for $n_2= 4,6...$ above 1.3 THz are well pronounced. The lowest odd mode $\omega_1(2)$ however, which lies at the frequency 0.9 THz (between $\omega_1(1) = 0.6$~THz and $\omega_1(3) = 1.3$~THz), can be barely seen near its crossings with the mode $\omega_2(1)$ at $w_2$ = 1.1~$\mu$m and mode $\omega_2(3)$ at $w_2$ = 3.3~$\mu$m. Therefore, the odd mode $\omega_1(2)$ is totally driven by the latter even modes which act as master oscillators. No anticrossing behavior is seen in this case because the odd mode provides only a weak feedback on the master (even) plasmon modes. The odd mode $\omega_2(2)$ is seen only for $w_2 < 1.5~\mu$m. Evidently, this occurs because the odd mode is more sensitive to the asymmetry of the 2D-PlCr unit cell in shorter plasmonic cavities where the electric field antinodes are closer to the edges of the conductive strip (for the same reason, the higher-order odd modes, having the antinodes close to the edges of the conductive strip, are well pronounced at high frequencies in the spectra shown in Fig. \ref{fateev5}). As also seen in Fig. \ref{fateev5}, the $n_2$ = 2 plasmon mode is resonantly excited driven by the $n_1$ = 1 plasmon mode at frequency 0.6 THz for $w_2 = 3.4~\mu$m.

\begin{figure}[ht]
\includegraphics{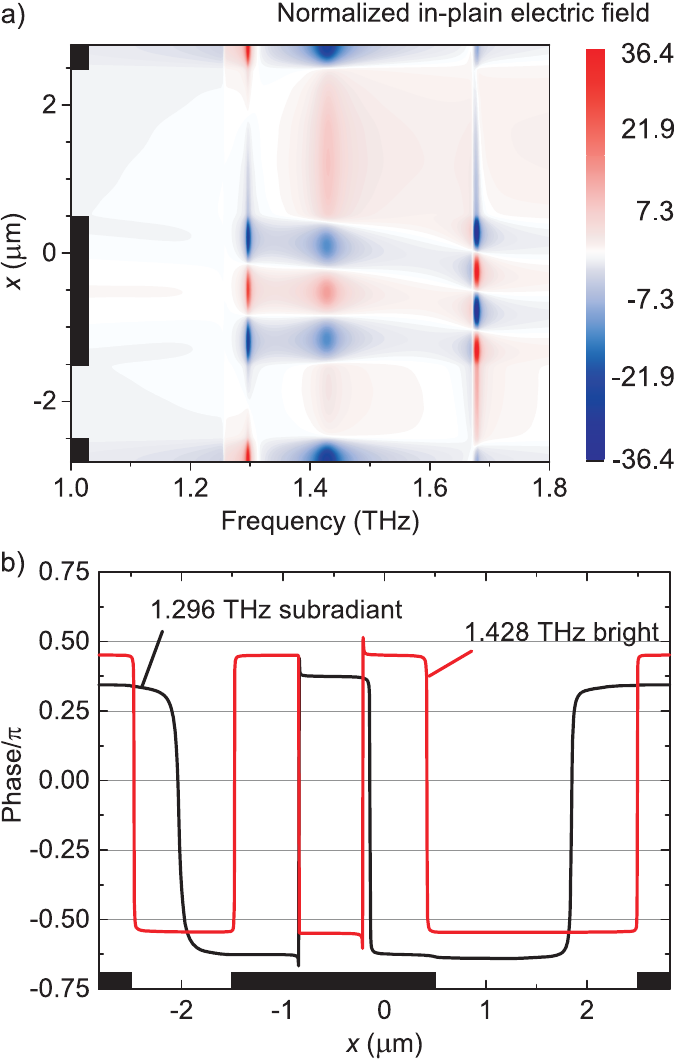}
\caption{(Color online) (a) The coupled plasmon mode profiles in the 2D-PlCr with an asymmetric unit cell at the anticrossing regime between the $n_1 = 3$ and $n_2 = 1$ plasmon modes at frequencies around 1.35 THz for $w_2 = 628$~nm and $\tau=7\times10^{-11}$~s. The in-plane electric field in 2D electron system is normalized to the electric-field amplitude in the incident THz wave. (b) The phases of the electric fields of the two different coupled plasmon modes as functions of the $x$-coordinate within the 2D-PlCr unit cell. The positions of the metal-grating strips are shown by thick black bars along the $x$-coordinate.}
\label{fateev7}
\end{figure}

Away from the anticrossing points, the normally incident THz radiation excites the plasmon modes with the electric field confined under the strips of either one or another sub-grating. Therefore, the profiles of the plasmon modes remain highly symmetric in the $x$-direction relative to the center of the respective strip and, hence, the amplitudes of the oppositely travelling Fourier harmonics remain equal in this case. In the region of anticrossing of two even modes (Fig. \ref{fateev7}), the plasmonic field of the coupled plasmon modes resides simultaneously under the conductive strips of 
both sub-gratings. Consequently, the total plasmonic field is asymmetric throughout the 2D-PlCr unit cell. However, in spite of the asymmetric profile of the total plasmonic field, the Fourier harmonics of this field are only slightly asymmetric, $ |\tilde{E}_{x,p}| \approx |\tilde{E}_{x,-p}|$, with the maximum relative difference between the amplitudes of the oppositely travelling Fourier harmonics of the electric field below $1\%$ and $10\%$ in the bright and subradiant modes, respectively. This is because the total plasmonic field of each coupled mode is a sum of the two fields resonantly excited under the two sub-gratings, while each of the fields is symmetric in its sub-grating and the phase shift between different antinodes of either plasmon mode is close to $\pi$ (see Fig. \ref{fateev7}(b)). This allows for approximating the total field as a product of a constant phase factor and a real function of $x$-coordinate in each coupled mode.

\begin{figure}[ht]
\includegraphics{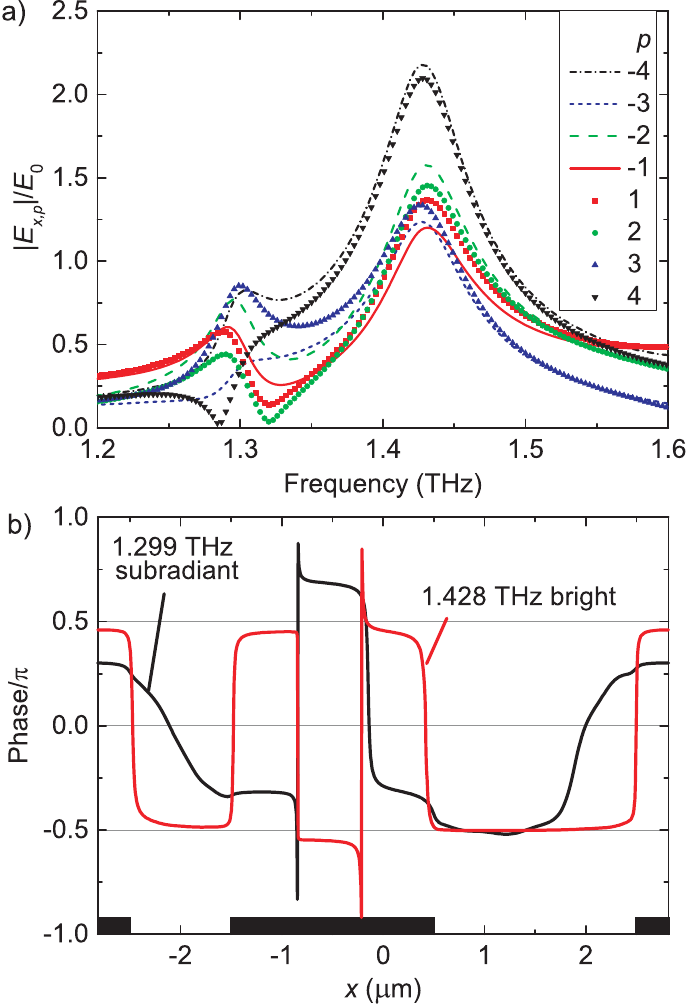}
\caption{ (Color online) (a) The spectra of different spatial Fourier harmonics of the plasmonic electric field in 2D electron system near the anticrossing regime between the $n_1 = 3$ and $n_2 = 1$ plasmon modes at frequencies around 1.35 THz for $w_2 = 628$~nm and $\tau=5\times10^{-12}$~s in the 2D-PlCr with an asymmetric unit cell. The amplitudes of the Fourier harmonics are normalized to the electric-field amplitude in the incident THz wave. (b) The phases of the electric field of the two different coupled plasmon modes as functions of the $x$-coordinate within the 2D-PlCr unit cell. The positions of the metal-grating strips are shown by thick black bars along the $x$-coordinate.}
\label{fateev9}
\end{figure}

\begin{figure}[ht]
\includegraphics{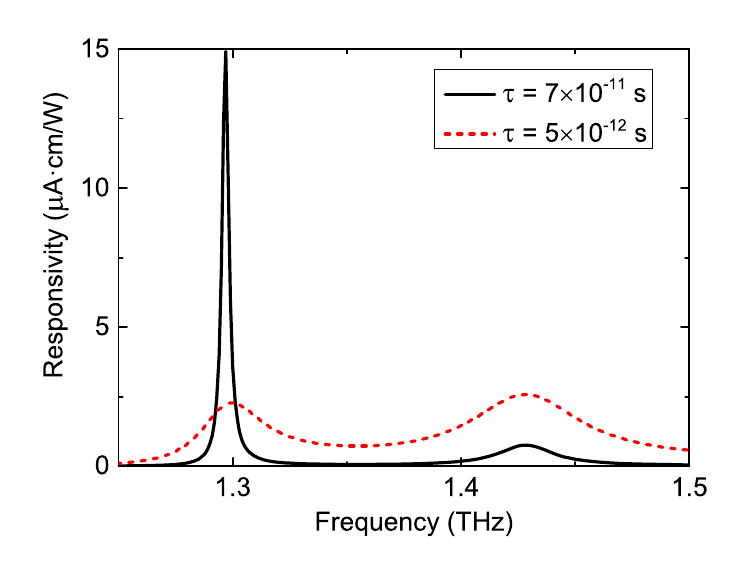}
\caption{(Color online) Responsivity peaks of the plasmon-photogalvanic current at the coupled resonances of the $n_1 = 3$ and $n_2 = 1$ plasmon modes at frequencies around 1.35 THz ($w_2 = 628$~nm) for two different values of the electron scattering time. The responsivity is defined as the ratio between the photocurrent density and the energy flux in the incident THz wave.}
\label{fateev12}
\end{figure}

A remarkable difference in the Fourier-harmonic amplitudes $ |\tilde{E}_{x,p}|$  and $ |\tilde{E}_{x,-p}|$  can be obtained for a shorter electron scattering time, which corresponds to the weak coupling regime. In this regime, the width of each plasmon resonance is comparable to (or larger than) the splitting of the coupled plasmon modes. As a result, 
 the plasmonic fields of the two coupled plasmon modes strongly overlap in the frequency domain which dramatically increases the asymmetry in the amplitudes of the oppositely travelling Fourier harmonics. In this case, the difference $ |\tilde{E}_{x,p}| - |\tilde{E}_{x,-p}|$  can become comparable with the amplitude of the Fourier-harmonic itself in the resonance of the subradiant plasmon mode (Fig. \ref{fateev9}). 
This is because the phase of the resonating coupled plasmon mode is disturbed due to its strong coupling with the off-resonance plasmon mode. It is seen from Fig. \ref{fateev9}(a) that the electric field at different antinodes of the subradiant plasmon mode oscillates strongly out of phase. The phase is predominantly disturbed in the subradiant plasmon mode due to large difference in the strength of the dipole moments of the subradiant and bright plasmon modes. Enhanced difference in the amplitudes of the oppositely travelling Fourier harmonics at the coupled plasmon resonances causes corresponding peaks in the photocurrent, Eq. (2), due to the differential plasmon drag (Fig. \ref{fateev12}). It is worth noting that, despite the asymmetry in the amplitudes of the oppositely travelling Fourier harmonics of the plasmon field is smaller for longer electron scattering time, the plasmon-photogalvanic current is much greater due to higher quality factor of the plasmon resonance in this case. 

\begin{figure}[ht]
\includegraphics{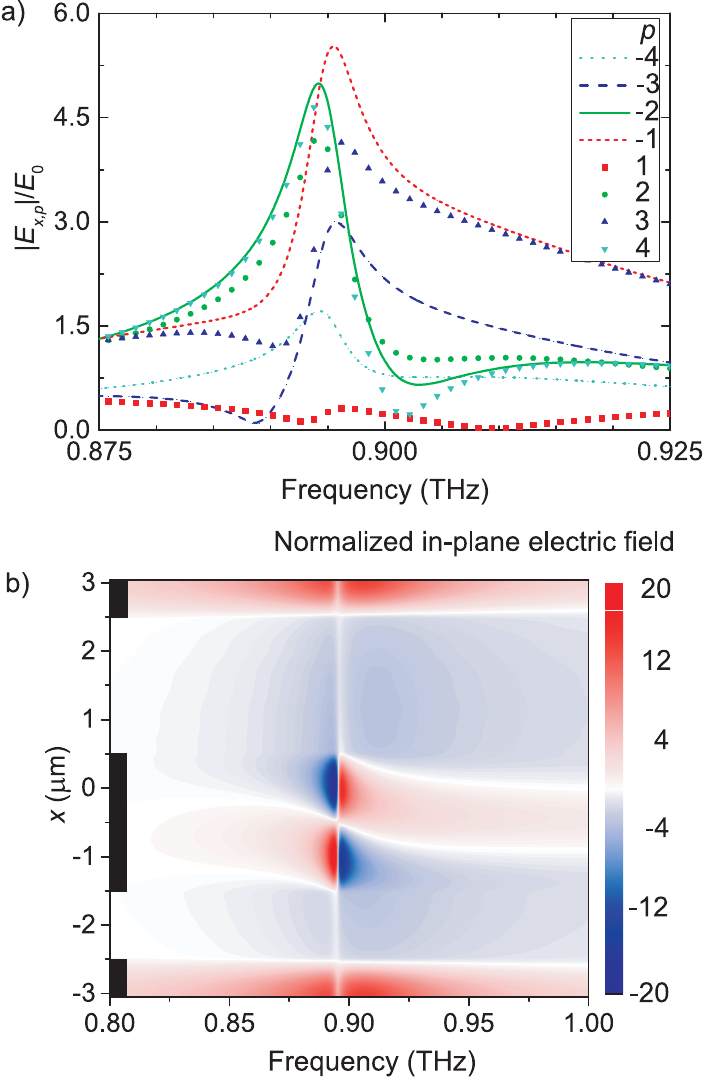}
\caption{(Color online) (a) The spectra of different spatial Fourier harmonics of the in-plane component of the plasmonic electric field in 2D electron system for the resonant excitation of the $n_1 = 2$ plasmon mode driven by the $n_2 = 1$ plasmon mode at frequencies around 0.9 THz for $w_2 = 1.1~\mu$m and the electron scattering time $\tau=7\times10^{-11}$~s  in the 2D-PlCr with an asymmetric unit cell. (b) The plasmon mode profile at the resonant excitation of the $n_1 = 2$ plasmon mode driven by the $n_2 = 1$ plasmon mode. The amplitudes of the Fourier harmonics and the in-plane electric field in 2D electron system are normalized to the electric-field amplitude in the incident THz wave. The positions of the metal-grating strips are shown by thick black bars along the $x$-coordinate in panel (b).}
\label{fateev10}
\end{figure}

\begin{figure}[ht]
\includegraphics{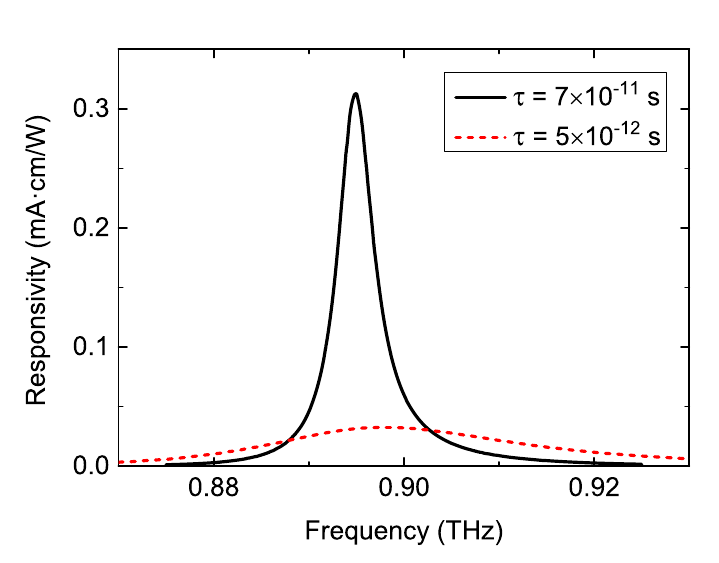}
\caption{(Color online) Responsivity peaks of the plasmon-photogalvanic current at the resonant excitation of the $n_1 = 2$ plasmon mode driven by the $n_2 = 1$ plasmon mode at frequencies around 0.9 THz ($w_2 = 1.1$~$\mu$m) for two different values of the electron scattering time.}
\label{fateev13}
\end{figure}

The asymmetry of the plasmonic field in the 2D-PlCr is even more enhanced in the frequency region of the resonant mixing of the even and odd plasmon modes excited under the conductive strips of different sub-gratings (Fig. \ref{fateev10}(a)). In this case, the difference in the amplitudes of the oppositely travelling Fourier harmonics is remarkable even for the longer electron scattering time because the electric fields of the two coupled modes strongly overlap in the frequency domain due to a small frequency splitting of these modes (Fig. \ref{fateev10}(b)). Correspondingly, the plasmon-photogalvanic current due to the differential plasmon drag becomes greater (Fig. \ref{fateev13}).

At the resonant frequencies, the electric field $|\tilde{E}_x(x)|$ in the 2D electron system is greater than the amplitude of the electric field in the incident THz wave roughly by the plasmon resonance quality factor which is typically 20-30 for  $\tau = 7 \times 10^{-11}$~s and 2-5 for $\tau = 5 \times 10^{-12}$~s at the frequency $\sim$~1~THz. Due to this fact, the plasmon-photogalvanic current in 2D electron system, Eq. (2), strongly enhances at the frequencies of the noncentrosymmetric plasmon resonances. Another reason for the strong plasmon-photogalvanic current is a large value of the plasmon wavevector which is from two to three orders of magnitude higher than the photon wavevector in THz frequency range. As a result, the plasmon-photogalvanic current exceeds the photon drag current in 2D electron system by several orders of magnitude.

\section{Conclusion}
In conclusion, we have shown that the plasmon-photogalvanic current can be excited in 2D-PlCr with an asymmetric unit cell via the differential plasmon drag of free carriers in 2D electron system by noncentrosymmetric plasmonic field. A strong asymmetry of the plasmon field arises due to the resonant coupling of different plasmon modes in the 2D-PlCr. Although we  limited ourselves to the analysis of the coupling and mixing between the screened plasmon modes, our results have a more general value because they provide a symmetry criteria which can also be applicable for building a strongly asymmetric plasmonic field in a polymodal 2D-PlCr supporting both screened and unscreened plasmon modes and incorporating a spatially periodic 2D electron plasma. Resonant excitation of the plasmonic field with a strong asymmetry gives a key to understanding the origin of ultra-strong THz plasmon-photogalvanic effects in 2D-PlCr.

\begin{acknowledgments}
VVP and DVF are gratefull to the Russian Foundation for Basic Research for support (15-02-02989). ELI thanks the Russian Science Foundation (14-12-01067). SDG acknowledges support from DFG (SFB 689). 
\end{acknowledgments}

\bibliography{fateev}

\end{document}